\documentclass[prl,twocolumn,floatfix,superscriptaddress]{revtex4-2}

\usepackage{amsmath, amssymb}
\usepackage{mathtools}
\usepackage{lipsum}
\usepackage{times}
\usepackage{graphicx}
\usepackage{wasysym}
\usepackage{bm}
\usepackage{hyperref}
\usepackage{xspace}
\usepackage{siunitx}
\usepackage{xcolor}
\usepackage{soul}
\usepackage{braket}

\definecolor{myColor}{rgb}{0.02,0.12,0.8}
\definecolor{myciteColor}{rgb}{0.39,1.0,0.89}
\hypersetup{colorlinks=true, linkcolor=myColor, filecolor=myColor, urlcolor=myColor,citecolor=myColor, urlcolor=myColor}
\graphicspath{{./Figures/}}

\DeclareSymbolFont{extraup}{U}{zavm}{m}{n}
\DeclareMathSymbol{\varheart}{\mathalpha}{extraup}{86}
\DeclareMathSymbol{\vardiamond}{\mathalpha}{extraup}{87}

\DeclareSIUnit{\nK}{\nano\kelvin}
\DeclareSIUnit{\aB}{\emph{a}_0}
\DeclareSIUnit{\G}{G}

\renewcommand{\figurename}[1]{Fig.~}

\newcommand\identity{1\kern-0.25em\text{l}}

\def\js#1{{\color{black} #1}}

\begin{document}

\title{Observation of a topological edge state stabilized by dissipation}
\author{Helene Wetter}
\affiliation{Physikalisches Institut, Universit\"at Bonn, Nussallee 12, 53115 Bonn, Germany}
\author{Michael Fleischhauer}
\affiliation{Department of Physics and Research Center OPTIMAS, RPTU Kaiserslautern-Landau, 67663 Kaiserslautern, Germany}
\author{Stefan Linden}
\affiliation{Physikalisches Institut, Universit\"at Bonn, Nussallee 12, 53115 Bonn, Germany}
\author{Julian Schmitt}
\email{schmitt@iap.uni-bonn.de}
\affiliation{Institut f\"ur Angewandte Physik, Universit\"at Bonn, Wegelerstr. 8, 53115 Bonn, Germany}

\date{\today}

\begin{abstract}
Robust states emerging at the boundary of a system constitute a hallmark for topological band structures. Other than in closed systems, topologically protected states can occur even in systems with a trivial band structure, if exposed to suitably modulated losses. Here, we study the dissipation-induced emergence of a topological band structure in a non-Hermitian one-dimensional lattice system, realized by arrays of plasmonic waveguides with tailored loss. We obtain direct evidence for a topological edge state that resides in the center of the band gap. By tuning dissipation and hopping, the formation and breakdown of an interface state between topologically distinct regions is demonstrated.
\end{abstract}

\pacs{03.65.Vf,42.82.Et,42.70.Qs}
\maketitle

Topology is an important paradigm for our understanding of phases of matter~\cite{Altland:1997}, with the quantum Hall effect constituting a prominent example of a topological system isolated from the environment~\cite{Klitzing:1980}. Interfacing materials with distinct topological properties has remarkable implications leading to localized edge states at the boundary, which due to their robustness against disorder are considered as valuable resource states for quantum technologies~\cite{Sarma:2015}. Conceptually, the robustness results from the existence of global integer-valued invariants, which can only change in a phase transition associated with the closing of a gap. Inspired by solid-state systems, topological states in closed Hermitian systems have been experimentally realized in a wide range of platforms, such as ultracold atoms or photonics~\cite{Leseleuc:2019,Ozawa:2019,Bleckmann:2017}.

Exploring topological phenomena in open systems presents a complementary approach to realize robust edge states, where the coupling between the system and the environment (\emph{e.g.} by pumping or dissipation of particles) acts as a resource rather than a limitation. Starting from the prediction of topological transitions in non-Hermitian quantum walk \cite{Rudner:2009}, conceptual questions about the classification of open-system topological phases for non-Hermitian and Lindbladian settings~\cite{Bardyn:2013,Linzner:2016,Gong:2018,Lieu:2020,Altland:2021}, the role of topological invariants and edge states~\cite{Schomerus:2013,Liang:2013,Leykam:2017,Yao:2018,Kunst:2018,Luo:2019}, and the band theory~\cite{Shen:2018,Yokomizo:2019} have been addressed theoretically. Experimentally, non-Hermitian systems have been realized in photonics, where driven-dissipative effects can be engineered~\cite{Feng:2017}. Combining topologically nontrivial photonic crystals with gain or loss, this has allowed for the observation of topological (lasing) states in waveguides~\cite{Zeuner:2015,Weimann:2017}, resonator arrays~\cite{Bahari:2017,Zhao:2018} and exciton-polaritons~\cite{St-Jean:2017,Klembt:2018}. Topological protection is here, however, inherited from the photonic band structure, and not from a coupling to reservoirs. The implementation of topological phases that solely arise from non-Hermiticity and lack a Hermitian counterpart, as proposed in refs.~\cite{Bardyn:2013,Linzner:2016,Takata:2018,Comaron:2020a} and realized with mechanical metamaterials~\cite{Fan:2022a}, acoustic cavities~\cite{Gao:2020,Gao:2021}, and electrical circuits~\cite{Liu:2020}, has so far remained elusive for optical systems.

\begin{figure}[t]
    \centering
    \includegraphics[width=1.0\columnwidth]{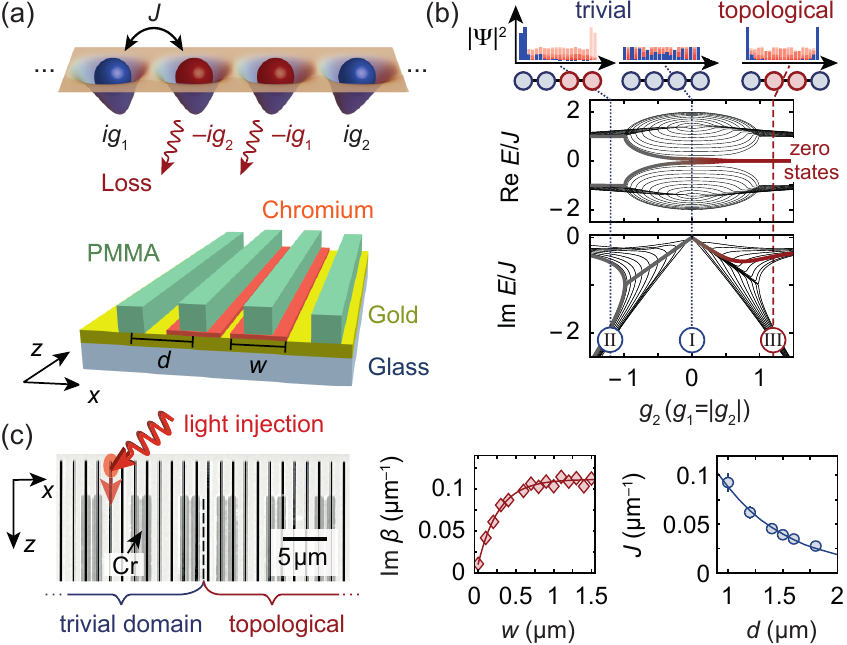}
    \caption{Experimental scheme. (a)~Lattice system with nontrivial topology induced solely by dissipation (top) and experimental realization with DLSPP waveguides spaced by distance $d$, where losses are induced by Chromium stripes of width $w$ (bottom). (b)~Complex-valued energy spectrum for 40 lattice sites with $g_1=|g_2|$. For $g_2<0$, the system is topologically trivial and the probability density $|\Psi|^2$ is concentrated in the first two lattice sites (top left). In the topologically nontrivial regime ($g_2>0$), $|\Psi|^2$ is exponentially localized at the edges (top right) and associated with midgap states at zero energy (red lines). (c)~{\itshape Left}: Waveguide sample with Chromium stripes (dark gray) arranged to realize trivial (topological) domains in the left (right) sample half. The Chromium-free region on top is used to excite the waveguides by grating coupling of laser light. {\itshape Right}: Measured dissipation in a single waveguide $\mathrm{Im}\beta$ versus $w$ and hopping $J$ between two waveguides versus $d$, along with fits (lines).}
    \label{fig:1}
\end{figure}

In this Letter, we report measurements of light-matter states with nontrivial topological properties, solely induced by tailored dissipation. Using surface plasmon polaritons (SPPs) confined in waveguide arrays, we obtain signatures for open-system topological edge states by identifying zero-energy modes localized at the boundary of the sample. The underlying one-dimensional (1D) lattice with 4-site unit cell realizes a non-Hermitian extension of the paradigmatic SSH model~\cite{Heeger:1988}, despite uniform hopping throughout the lattice. By tuning dissipation and hopping, the birth and death of a non-Hermitian topologically protected edge state is demonstrated.

The basic principle of our topological system, see Fig.~\ref{fig:1}(a), relies on a 1D lattice with spatially uniform nearest-neighbor hopping $J$ and spatially varying dissipation at the lattice sites~\cite{Takata:2018,Comaron:2020a}. The unit cell consists of 4 sites spaced by $d$, which are subject to a gain-loss pattern $(ig_1,-ig_2,-ig_1,ig_2)$ with real-valued dimensionless amplitudes $g_{1,2}$. The Bloch Hamiltonian of the open system at wave vector $k_x$
\begin{equation}
\hat H_{k_x} = J
    \begin{pmatrix}
ig_{1} & 1 & 0 &  e^{-4 i  k_x d} \\
1 & -ig_{2} & 1 & 0 \\
0  & 1  & -ig_1 & 1  \\
 e^{4 i k_x d} & 0 & 1 & ig_2
\end{pmatrix} - i J g_0 \identity
\label{eq:1}
\end{equation}
is a non-Hermitian matrix, \emph{i.e.}, $\hat H\neq \hat H^\dagger$, with complex energy eigenvalues $E$. Here, $g_0>0$ accounts for a loss at all lattice sites, which governs our implementation using purely dissipative waveguides, see Fig.~\ref{fig:1}(a) (bottom). Due to the global loss the steady-state is here the trivial vacuum. Nevertheless, as we will demonstrate in our experiments, the system possesses a nontrivial \emph{non-equilibrium} topology, protected by the symmetries of the Liouvillian $\cal L$ that governs the dynamics of the density matrix $\rho$, according to $\dot \rho = {\cal L} \rho$~\cite{Altland:2021}. The dynamical generators fulfill time reversal ($\sf T$), charge-conjugation ($\sf C$), and chiral ($\sf S$) symmetries~\footnote{See Supplemental Information.}, characterizing the class BDI~\cite{Altland:1997}, which can have a nontrivial topology in 1D.

The topological character of the 4-site model becomes apparent when considering the energy band structure and eigenstates in a finite-length lattice for different loss patterns, as shown in Fig.~\ref{fig:1}(b); for simplicity, we consider the symmetric case $g_1=|g_2|$ and $g_0=|g_2|$ with $g_2$ as the control parameter. In the different parameter regimes, $\mathrm{diag}(\hat H)=(0,0,0,0)$ for $g_2=0$ (lossless trivial, I), $\mathrm{diag}(\hat H)=-2iJ|g_2|(0,0,1,1)$ for $g_2< 0$ (dissipative trivial, II), and $\mathrm{diag}(\hat H)=-2 i Jg_2(0,1,1,0)$ for $g_2>0$ (topologically nontrivial, III), respectively (blue and red circles in Fig.~\ref{fig:1}(b)). Phase (I) exhibits a metal-like gapless band structure with probability densities $|\Psi|^2$ delocalized in the bulk, while the band structure in phase (II) is gapped. In striking contrast, phase (III) features two midgap states at zero energy in $\mathrm{Re}E$ (red lines), which are localized at the boundary of the system and decay exponentially into the bulk. Due to the dissipative nature of the system, $\mathrm{Im}\,E<0$ for all states. Note that while there is not yet a general understanding of non-equilibrium invariants for density matrices, a topological number for the non-Hermitian system has been identified theoretically as \js{a global Berry phase, which takes discrete values $W=0$ or $1$ for $g_1g_2<0$ or $g_1g_2>0$ in phases (II) or (III), respectively~\cite{Liang:2013,Takata:2018}; for details, see~\cite{Note1}.} Based on the non-equilibrium symmetries~\cite{Note1} and the above mentioned spectral and spatial signatures, topological states solely induced by dissipation rather than Hermitian band engineering are theoretically expected in our system.

\begin{figure}[t]
    \centering
    \includegraphics[width=1.0\columnwidth]{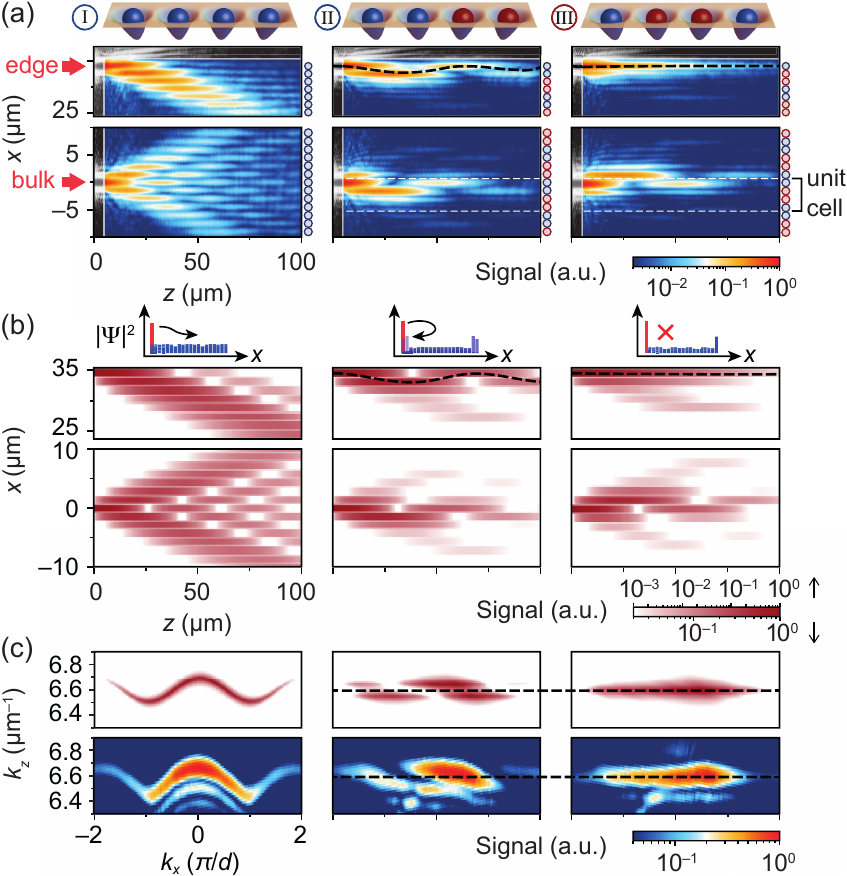}
    \caption{SPP evolution for metal-like (left, I), dissipative topologically trivial (middle, II) and dissipation-induced topological domains (right, III). (a)~Real-space intensity distribution for injection at the edge (top) and at the first site of a unit cell in the bulk (bottom), for a sample of 48 waveguides spaced by $d=\SI{1.4}{\micro\meter}$, corresponding to $J=0.045(3)\SI{}{\micro\meter}^{-1}$. Solid and white dashed lines indicate sample boundaries and unit cells\js{, and black dashed lines give the center-of-mass.} For the used $w=\SI{0.7}{\micro\meter}$ in (II) and (III), $\mathrm{Im}\beta=\SI{0.1}{\micro\meter}^{-1}$ and $g_1=|g_2|=1.1$. (b)~Simulated intensity evolution in the arrays from (a) with $J=\SI{0.045}{\per\micro\meter}$ and $\text{Im}\beta=\SI{0.1}{\per\micro\meter}$ (\SI{0.01}{\per\micro\meter}) for the lossy [red in (a)] and low-loss [blue in (a)] waveguides, in good agreement with the experimental results. (c)~Momentum-resolved energy spectra with $\cos$-shaped band for bulk excitation, band gap in the trivial and flat band in the topological domains, respectively, for edge excitation~\cite{Note1}, along with theory prediction (top). \js{Dashed lines indicate the energy of the topological zero mode (right), which lies between the bulk bands of case (II) (middle).}
    }    
    \label{fig:2}
\end{figure}

To experimentally investigate the topological properties of the non-Hermitian lattice system, we utilize SPPs confined in evanescently coupled arrays of dielectric loaded SPP waveguides (DLSPPWs) with tailored losses. The samples are fabricated by two-step electron beam lithography~\cite{Note1,Fedorova:2021}. Figure~\ref{fig:1}(a) (bottom) outlines a typical waveguide structure, realized by depositing PMMA ridges of about 200nm width and center-to-center spacing $d$ on top of a glass substrate, previously coated with a low-absorption 60nm-thin Gold layer to host the plasmonic part of the polaritons. Losses at individual lattice sites are introduced and controlled by adding Chromium stripes of variable width $w$ below the ridges.

An exemplary sample containing two different loss patterns in the unit cells, corresponding to the topologically trivial (II) and nontrivial (III) region, is shown in Fig.~\ref{fig:1}(c). The SPP evolution in the array is excited with a 980nm laser at \SI{15}{\micro\watt} optical power and characterized by leakage radiation microscopy~\cite{Bleckmann:2017,Note1}. By varying $w$ and $d$, both the additional absorption $\mathrm{Im}\beta$ from Chromium and the hopping $J$ (both in units $\SI{}{\micro\meter}^{-1}$) can be accurately controlled, see Fig.~\ref{fig:1}(c) (right); here $\beta$ denotes the complex-valued propagation constant in a single waveguide~\cite{Bleckmann:2017}. In terms of eq.~\eqref{eq:1}, the dissipation parameters follow as $g_{2}=\mathrm{Im}\beta/(2J)$. 

First, we study the evolution of the SPPs upon injecting a wave packet at the edge and in the bulk of the waveguide arrays, respectively, for three loss patterns according to phases (I) (metal-like), (II) (dissipative trivial), and (III) (dissipative topological). Figure~\ref{fig:2}(a) shows the real-space SPP intensity distributions obtained by imaging the leakage radiation~\cite{Note1}. For phase (I) (Fig.~\ref{fig:2}(a), left column), the SPP evolution mimics a two-state quantum walk of a particle in a periodic potential. This metal-like behaviour is highlighted by a conical transport along with a characteristic interference pattern when injecting the wave packet in the bulk; upon injection at the edge, the wave packet simply propagates in the $-x$ direction. With losses as in the topologically trivial phase (II) (Fig.~\ref{fig:2}(a), middle), the \js{ballistic transport} is inhibited and an oscillation of the intensity between two neighboring low-loss waveguides is observed for both excitation protocols. In contrast, with losses as in the topologically nontrivial phase (III) (Fig.~\ref{fig:2}(a), right), the edge excitation reveals apart from the overall damping a quasi-stationary intensity evolution that remains locked to the outermost waveguide. This localization occurs at the edge only, as understood from the corresponding probability density $|\Psi|^2$ shown in Fig.~\ref{fig:1}(b) (top), while excitation in the bulk reveals the same oscillatory dynamics as in phase (II), except for a phase shift due to the neighboring low-loss waveguide now lying above the excited one. \js{The oscillation results from a beating between bonding and antibonding states in the hybridized neighboring lattice sites [Fig.~\ref{fig:2}(c)]. In contrast, the absence of a beating indicates the zero energy of the topological edge state.} The phenomenology observed in the measured real-space intensity is in good agreement with numerical simulations based on coupled mode theory~\cite{Note1}, see Fig.~\ref{fig:2}(b), giving conclusive evidence that the non-Hermitian model is well-captured by our DLSPPW platform.

\begin{figure}[t]
    \centering
    \includegraphics[width=1.0\columnwidth]{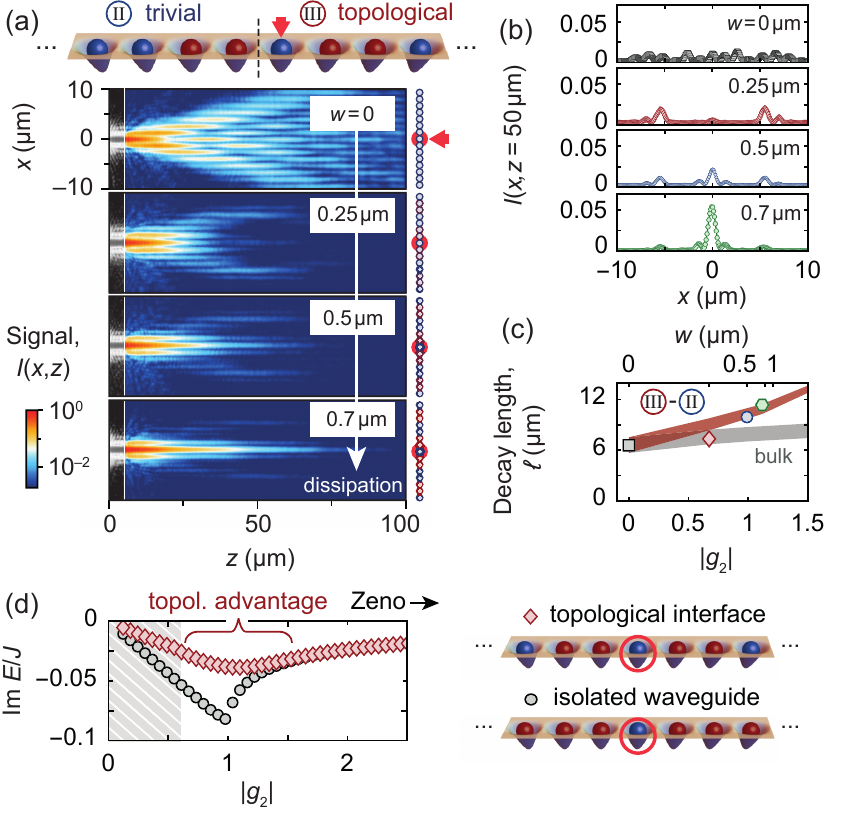}
    \caption{Dissipation-induced emergence of interface state between distinct topological domains. (a)~Interface between phase-(II) and phase-(III) domains realized by different loss patterns, and real-space intensity distribution $I(x,z)$ upon excitation of the interface waveguide (red) for increasing losses $\mathrm{Im} \beta=\{0,0.06,0.09,0.1\}\SI{}{\micro\meter}^{-1}$ from larger Chromium widths $w$. Losses correspond to $g_1=|g_2|=\{0,0.7,1,1.1\}$ and, as before, $J=0.045(3)\SI{}{\micro\meter}^{-1}$. \js{(b)~Line profiles of the intensity averaged around $z=\SI{50}{\micro\meter}$. (c)~Decay length $\ell$ of intensity in interface waveguide for cases in (a); for fitted line profiles $I(x=0,z)$, see~\cite{Note1}.} Despite increased Chromium absorption, an enhanced propagation distance is observed, providing evidence for a topologically robust state. Shaded areas show simulations for the interface (red) and bulk (gray), where the enhanced $\ell$ indicates the interface robustness beyond $g_{2}\approx 0.7$. \js{(d) Advantage of topological interface over isolated waveguide, seen in reduction of calculated $\mathrm{Im}\,E$ when induced losses and hopping are of the same order $g_2\sim 1$. For small $g_2$ (shaded) the edge state delocalizes, for large $g_2$ Zeno-like decoupling dominates.}
    }    
    \label{fig:3}
\end{figure}

Figure~\ref{fig:2}(c) shows the momentum-resolved occupation of the energy bands within the first two Brillouin zones from $k_x=-2\pi/d$ to $2\pi/d$, as obtained by recording the leakage radiation in the back-focal Fourier plane of the microscope objective\js{~\cite{Note1}}. In the metal-like phase (I) and for bulk excitation, the spectrum matches the expected $\cos$-shaped energy band that complements the independently observed ballistic transport in real space. The spectrum agrees with the simulated one in Fig.~\ref{fig:2}(c), except for a circular segment at $k_z\lesssim\SI{6.5}{\micro\meter}^{-1}$ that is well understood to arise from unconfined SPP propagation outside of the array~\cite{Drezet:2008}, also visible in the gray shaded area in Fig.~\ref{fig:2}(a). In the dissipative topologically trivial phase (II) and for edge excitation, the momentum distribution considerably changes. Two energy bands separated by a gap near $k_z=6.56(2)~\SI{}{\micro\meter}^{-1}$ and visible at $k_x\approx -0.5\pi/d$ are observed. In the topologically nontrivial phase (III), on the other hand, the momentum distribution exhibits only a single flat energy band centered at $k_z=6.59(2)~\SI{}{\micro\meter}^{-1}$, with a spectral width determined by losses and residual transport in the bulk. By comparing with Fig.~\ref{fig:1}(b), the data gives evidence for a topological zero-state in the band gap.

A unique feature of the investigated system lies in the fact that topological properties emerge as a consequence of dissipation alone, in a lattice which would be otherwise topologically trivial. To systematically test this dissipation-induced birth of topological order, we focus on an interface between two distinct domains prepared in phases (II) and (III), respectively, and gradually increase the loss $\mathrm{Im}\beta$. Figure~\ref{fig:3}(a) shows the real-space SPP evolution for increasing Chromium widths $w$, after consistently exciting the same low-loss waveguide, which is located at the interface. The conical intensity spread into the bulk for $w=0$, previously seen in Fig.~\ref{fig:2}, is gradually transformed into a quasi-stationary, \emph{i.e.}, transversely localized occupation of the interface waveguide. Remarkably, despite larger absorption $\mathrm{Im}\beta$, the SPP propagation length is significantly enhanced. This is understood from the increase of $\mathrm{Im}\,E$ for the topological zero-states for large enough $g_2\gtrsim 0.7$ [see Fig.~\ref{fig:1}(b) (bottom panel, red line)], marking a clear distinction point from its more lossy topologically trivial counterpart in a phase-II system for $g_2<0$ [Fig.~\ref{fig:1}(b) (bottom panel, gray line)]. Thus, the intensity becomes more localized and long-lived at the interface due to the presence of topologically distinct domains. The extended propagation distance is visually more striking in the line profiles at $z=\SI{50}{\micro\meter}$ in Fig.~\ref{fig:3}(b). Quantitatively, Fig.~\ref{fig:3}(c) shows the fitted $1/e$ decay length of the intensity at the interface as a function of $w$, along with the theoretically expected decay length at the interface waveguide and in the bulk, confirming the genuine dissipation-enhanced topological robustness of the interface state. \js{To emphasize the effect of topology, Fig.~\ref{fig:3}(d) shows calculations in which we compare $\mathrm{Im}\,E$ of the topological interface state to that of a mode localized at an isolated low-loss waveguide embedded in a bulk of lossy waveguides. In an intermediate regime where losses and hopping are of the same order, $0.6\lesssim g_2\lesssim 1.5$, a smaller absorption is observed for the topological configuration. This means that light transport along an interface between distinct topological domains is indeed expected to be enhanced. For larger values of $g_2$, Zeno-like decoupling dominates and both configurations show essentially the same losses.}

\begin{figure}[t]
    \centering
    \includegraphics[width=1.0\columnwidth]{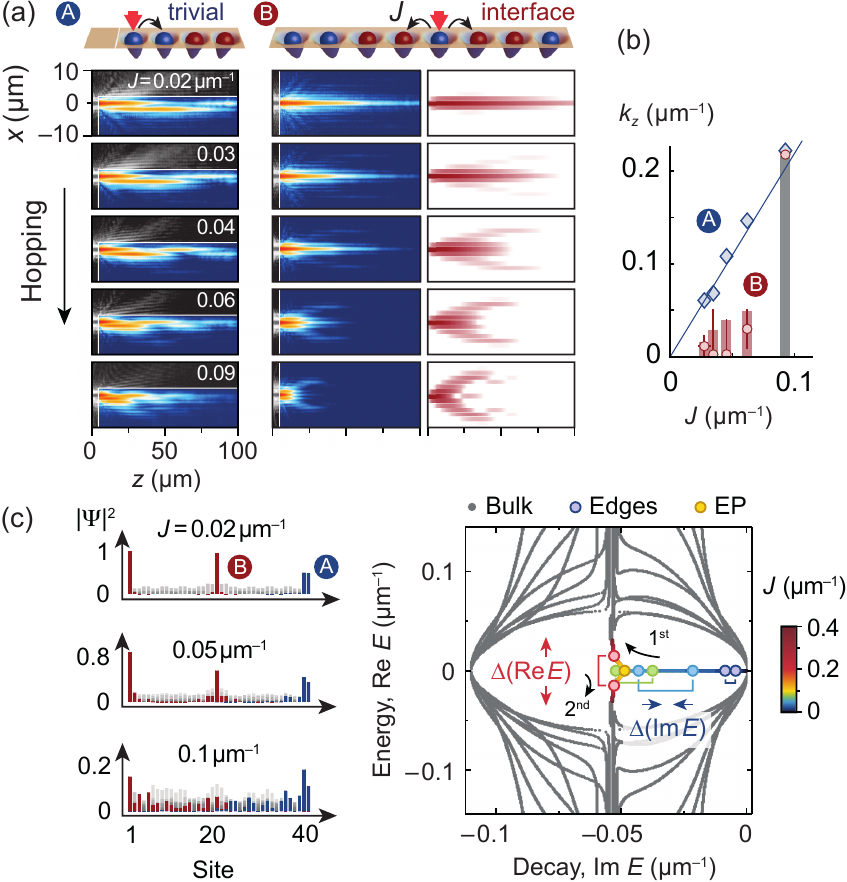}
    \caption{Breaking the topological protection at an exceptional point. (a)~\js{SPP evolution for increasing hopping $J$, realized by reduced waveguide spacings $d=\{1.8, 1.6, 1.4, 1.2, 1.0\} \SI{}{\micro\meter}$ at $w=\SI{0.7}{\micro\meter}$ ($\mathrm{Im}\beta=\SI{0.1}{\micro\meter}^{-1}$). For excitation at the trivial edge (A), the population oscillates faster as $J$ increases. Upon injecting light into the (II)-(III) interface (B), a quasi-stationary edge state is observed, which at $J\approx \SI{0.09}{\micro\meter}^{-1}$ becomes delocalized in the bulk, in agreement with simulations (right). (b)~Oscillation frequency $k_z$ at excited waveguide fitted with $I(z)=a_1\cos(k_z z+\phi )e^{-z/\ell}+a_0$. Solid line and bars give numerical results, and error bars denote uncertainties of fit parameter. (c)~{\itshape Left:} Calculated probability density at boundaries,} along with bulk modes (gray). A delocalization of the interface mode (red) is visible for too large $J$, indicating the loss of topological character. {\itshape Right:} Complex eigenenergies of the two topological (colored) and 38 bulk (gray) states, as $J$ is varied. For $J\approx\SI{0.095}{\micro\meter}^{-1}$, the eigenvalues coalesce at an exceptional point (yellow).} 
    \label{fig:4}
\end{figure}

Conversely to the discussed formation of an edge state upon introducing dissipation, we next focus on \emph{breaking} the topological protection at the interface by increasing the hopping $J$ in the presence of dissipation. For this, we have reduced the waveguide spacing $d$, while keeping the losses in the interfaced phases (II) and (III) fixed at $\mathrm{Im}\beta=\SI{0.1}{\micro\meter}^{-1}$ [with $w=\SI{0.7}{\micro\meter}$ as in Fig.~\ref{fig:2}]. Figure~\ref{fig:4}(a) shows the real-space SPP evolution after injection of a wave packet at the boundary between a trivial phase-(II) domain and the vacuum. Not surprisingly, the population oscillates between the two low-loss waveguides as before, but now with a larger wave vector $k_z$ as $J$ is increased; see the blue data in Fig.~\ref{fig:4}(b). At the phase (II)-(III) interface shown in Fig.~\ref{fig:4}(a), the SPP population stays localized at the interface waveguide without any appreciable transverse transport or oscillation; see the red data in Fig.~\ref{fig:4}(b). Eventually, at $J\approx \SI{0.09}{\micro\meter}^{-1}$, the tunnel coupling becomes so large that the SPPs propagate away from the interface into the bulk, breaking up the topological edge state.
 
The diminished topological protection is understood to result from a delocalization of the interface mode with increased $J$, as shown in Fig.~\ref{fig:4}(c). The calculated $|\Psi|^2$ is shown for the trivial edge and interface modes, along with bulk modes. For larger $J$, we find the transverse localization length of the interface mode to extend further, revealing an exponential decay of $|\Psi|^2$ into the bulk with a maximum probability that consistently occurs at the interface waveguide. For the trivial edge, however, the maximum of the probability soon shifts away from the boundary of the sample and generally does not exhibit an exponential decay towards the bulk region. This phenomenology shares a close analogy with the topological edge states encountered in the SSH model~\cite{Heeger:1988}, where the probability density of the dimerized edge modes decays exponentially. In the non-Hermitian 4-site model, the alternating hopping of the SSH model is replaced by an effective (de-) coupling of pairs of gain-gain or loss-loss (mixed gain-loss) lattice sites~\cite{Takata:2018}, and a dimerization with two- and four-site periodicity occurs in $\mathrm{Re}\Psi$ and $\mathrm{Im}\Psi$, respectively. 

Finally, the spectrum of the complex energy eigenvalues in Fig.~\ref{fig:4}(c) provides a physical picture about the broken topological protection of the interface state (colored circles; energies of bulk modes are shown in gray) near $J\approx\SI{0.095}{\micro\meter}^{-1}$. For small $J$, the eigenvalues of the zero-states with $\mathrm{Re}E=0$ fall in the band gap and are separated in $\mathrm{Im}\,E$ (blue, green circles); note that due to $\mathrm{Im}\,E\neq 0$ the non-Hermitian system is not $\mathcal{P}\mathcal{T}$-symmetric~\cite{Feng:2017}. As $J$ increases, the imaginary gap closes and the eigenvalues coalesce at an exceptional point (yellow circle), followed by an opening of a real energy gap, which lifts the edge state degeneracy and merges both with the bulk bands, visible in Fig.~\ref{fig:4}(c) (red circle). At the exceptional point, the system has lost its topological character.

In conclusion, we have experimentally demonstrated open-system topological states induced by dissipation alone, using SPP waveguide arrays with uniform hopping and spatially-distributed loss. Evidence for the topological nature of the non-Hermitian system is obtained from a localized midgap edge state between distinct topological domains. By independently tuning dissipation and hopping, both the emergence and breaking of topological order is observed. For the future, lowering the SPP losses may enable direct measurements of the topological invariant by interferometry~\cite{Longhi:2013} and give access to non-Hermitian Floquet engineering by modulated loss and hopping~\cite{Fedorova:2020,Fedorova:2021}. An intriguing perspective lies in the implementation of open-system topological states with optical quantum gases within optically-active microcavities~\cite{Kurtscheid:2020,Busley:2022}, opening ways to study the fate of topological order in the presence of fluctuations in one and two dimensions~\cite{Schmitt:2014,Luo:2019}. \js{In combination with gain, dissipation-induced topological edge states are also interesting candidates for applications, such as robust routing of light in reconfigurable interface channels, \emph{e.g.}, for optical interconnects.}

We thank V. Zimmermann and Z. Fedorova for discussions. S.L., M.F. and J.S. acknowledge support from the DFG within SFB/TR 185 (277625399). J.S. acknowledges support by the EU (ERC, TopoGrand, 101040409), and by DFG within the Cluster of Excellence ML4Q (EXC 2004/1–390534769).

\bibliography{references}




\section{Supplementary Material}

\setcounter{figure}{0}
\renewcommand{\thetable}{S\arabic{table}}
\renewcommand\thefigure{S\arabic{figure}}
\renewcommand{\theHtable}{Supplement.\thetable}
\renewcommand{\theHfigure}{Supplement.\thefigure}

\section{Sample preparation}
Our waveguide samples are fabricated in a multi-step process using electron beam lithography (EBL), as described in ref.~\cite{Fedorova:2021}. In a first step, Poly(methyl methacrylate) (PMMA) is used as a positive tone resist at an electron dose of \SI{2.45}{\coulomb\per\square\meter} to create a mask for the \SI{6}{\nano\meter} chromium layer thermally evaporated on top. A lift-off process reveals the chromium stripes, which introduce the losses in the dissipative waveguide samples. In the second EBL step, the waveguides are created by exposing the PMMA to an electron dose of \SI{7}{\coulomb\per\square\meter}; here the PMMA acts a negative tone resist. The width of each waveguide is approximately \SI{200}{\nano\meter}. A scanning electron micrograph of an exemplary waveguide sample equipped with Chromium stripes is shown in Fig.~1(c) (left) of the main text. \js{By varying the width of the Chromium stripes $w$ and the spacing of the waveguides $d$, the absorption $\mathrm{Im}\beta$ and the hopping $J$ are controlled. The corresponding data shown in Fig.~1(c) (right) of the main text has been obtained by measuring the decay of the intensity in a single waveguide and by recording the coherent oscillation between two adjacent waveguides in separate samples, respectively.}

\section {Experimental system}
\begin{figure}[t]
    \centering
    \includegraphics[width=1.0\columnwidth]{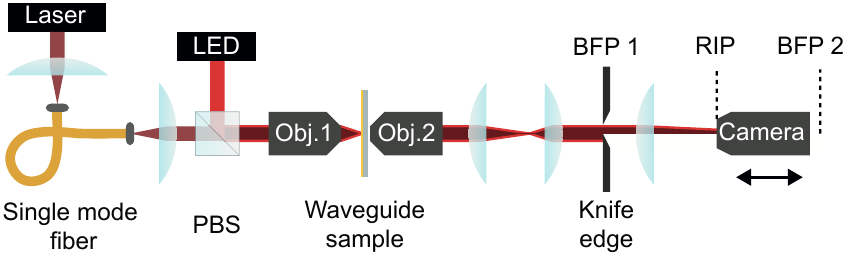}
\caption{\js{Sketch of the experimental setup. The laser beam used for the waveguide excitation is coupled to a single-mode fiber, TM-polarized by a polarizing beam splitter (PBS) and focused onto the sample using a microscope objective (Obj. 1). The optional LED light source provides background illumination for alignment of the sample. The leakage radiation is collected by a second microscope objective (Obj. 2) and imaged onto the camera placed in either the real-space image plane (RIP) or the second back-focal plane (BFP 2) to give the the real-space and Fourier space intensity distributions, respectively. The knife edge placed in the first back-focal plane (BFP 1) is used to filter out residual light from the transmitted laser beam.}}
    \label{fig:Sl1}
\end{figure}

\js{A schematic of the experimental setup is shown in Fig.~\ref{fig:Sl1}}. The surface plasmon polaritons (SPPs) are excited in a site-resolved way by focusing a TM-polarized cw laser beam at \SI{980}{\nano\meter} wavelength and \SI{15}{\micro\watt} optical power onto a grating coupler placed on top of one waveguide, which diffracts the laser light into the targeted waveguide. \js{In addition, an LED source is used for alignment and positioning of the waveguide sample. Before exciting the waveguides, the transverse laser beam profile is cleaned by coupling the laser to a single-mode fiber. The polarization of the laser is controlled by a polarizing beam splitter cube (PBS).} 

\js{The leakage radiation emitted by the SPPs is collected by an oil-immersion objective ($\times 60$ magnification, NA$=1.4$, Objective~2 in Fig.~\ref{fig:Sl1}) and imaged onto an sCMOS camera (Andor Zyla) positioned in the real image plane (RIP) to obtain the real-space intensity distribution. Leakage radiation microscopy also enables recording the momentum-resolved Fourier space with a single measurement by imaging the back-focal plane (BFP) of the microscope objective. For this purpose, residual light of the transmitted laser beam is filtered out by a knife edge in the intermediate first back-focal plane of the oil immersion objective. The image taken by the camera positioned in the second BFP yields the Fourier transform of the two-dimensional real-space field distribution, which is not accessible by analytically Fourier transforming the real-space intensity data. The corresponding spectra are a function of longitudinal and transverse momenta $k_z$ and $k_x$, respectively, and are shown in Fig.~2(c) of the main text.}

\js{In the measured spectral distribution of the lossless phase (I), see Fig.~2(c) of the main text (bottom left panel), residual emission at values of $k_z$ below the $\cos$-shaped bulk band is visible. The signal is attributed to arise from leakage radiation of free SPPs propagating outside of the waveguide array, as has been characterized in previous work~\cite{Drezet:2008}. Because of the unequal aspect ratio of the plotted $k_x$ and $k_z$ axes, the actually circularly symmetric propagation appears elliptic and only the top segment is visible within the shown plot range. Beyond the first Brillouin zone $|k_x|>\pi/d$ the signal vanishes, indicating that it is not a replica of the $\cos$-shaped band.}

\section{Symmetry classification of topology}
Generalized unitary and anti-unitary symmetries are a powerful tool to classify topological properties of systems of non-interacting particles. The ten fundamental symmetry classes of Hamiltonians~\cite{Altland:1997} have lead to the famous periodic table of topological insulators and superconductors \cite{Ryu:2010,Kitaev:2009}. Out of thermal equilibrium, a symmetry classification must include besides the Hamiltonian also the generators describing dissipation and the associated fluctuations~\cite{Altland:2021,Lieu:2020}. If the non-equilibrium dynamics is described by a Lindblad master equation for the density operator $\rho$
\begin{equation}
    \partial_t\rho = -i[\hat H,\rho] +\sum_\mu\Bigl( 2 \hat L_\mu \rho \hat L_\mu^\dagger - \{\hat L_\mu^\dagger \hat L_\mu,\rho\}\Bigr),
\end{equation}
where the Hamiltonian $\hat H$ is bilinear in particle creation and annihilation operators and the jump operators $\hat L_\mu$ are linear, a complete classification of generalized symmetries can be given in terms of $\hat H$ and the Hermitian operator
\begin{equation}
    \hat M = \sum_\mu \hat L_\mu^\dagger \hat L_\mu.
\end{equation}
We here follow the approach of Ref.~\cite{Altland:2021} where non-interacting fermions have been considered. Although the SPPs in our experiment are bosons, we can directly apply the explicit formulation in terms of free fermions since the physics in our non-interacting system is governed by single-particle physics only. To this end we write $\hat H = \frac{1}{2} \hat A^\dagger\,  {\sf H} \hat A$ and $\hat M = \frac{1}{2} \hat A^\dagger\, {\sf M} \hat A$ in a symmetric form, where $\hat A =(\hat c_1,\hat c_1^\dagger, \hat c_2,\hat c_2^\dagger,\dots )^\top$ is a vector of fermion annihilation and creation operators $\hat c_i, \hat c_i^\dagger$ at lattice site $i$, respectively, with $\sf H$ and $\sf M$ being Hermitian matrices. Furthermore, $\sf M$ is decomposed as
\begin{align}
    {\sf M} = {\sf D}- & i{\sf P},\\
    {\sf D} = \frac{1}{2}\left({\sf M} +\sigma_x {\sf M}^\top \sigma_x\right),\quad & 
    {\sf P} = \frac{i}{2}\left({\sf M} -\sigma_x {\sf M}^\top \sigma_x\right),
\end{align}
where the Pauli matrix $\sigma_x$ corresponds to the exchange of annihilation and creation operators in $\hat M$.

In the experimental system the stationary state is the vacuum, and the presence of edge states is solely a consequence of \emph{non-equilibrium} topology. As shown in Ref.~\cite{Altland:2021}, the generator of the non-equilibrium dynamics of a fermion lattice model is invariant under time reversal ${\sf T}$ if a unitary matrix ${\sf U}_T$ exists such that in Bloch-momentum space with wave number $k$ the following relations are fulfilled: 
\begin{align}
    {\sf U}_T \, {\sf H}^*(-k) \, {\sf U}_T^{-1} &=  - {\sf H}(k),\nonumber \\
    {\sf U}_T \, {\sf D}^*(-k) \, {\sf U}_T^{-1} &=  + {\sf D}(k),\label{eq:T}\\
    {\sf U}_T \, {\sf P}^*(-k) \, {\sf U}_T^{-1} &=  - {\sf P}(k).\nonumber
\end{align}
Note the minus sign in the first equation, which is different from the equilibrium case. Similarly, one finds that the non-equilibrium dynamics is invariant under charge conjugation $\sf C$ if a unitary matrix ${\sf U}_C$ exists, such that 
\begin{align}
    {\sf U}_C \, {\sf H}^\top(-k) \, {\sf U}_C^{-1} &=  - {\sf H}(k),\nonumber \\
    {\sf U}_C \, {\sf D}^\top(-k) \, {\sf U}_C^{-1} &=   + {\sf D}(k),\label{eq:C}\\
    {\sf U}_C \, {\sf P}^\top(-k) \, {\sf U}_C^{-1} &=  - {\sf P}(k).\nonumber
\end{align}
Finally, there is chiral symmetry $\sf S = \sf T \cdot \sf C$, if 
\begin{align}
    {\sf U}_S \, {\sf H}^\dagger(k) \, {\sf U}_S^{-1} &=  + {\sf H}(k),\nonumber \\
    {\sf U}_S \, {\sf D}^\dagger(k) \, {\sf U}_S^{-1} &=  + {\sf D}(k),\label{eq:S}\\
    {\sf U}_S \, {\sf P}^\dagger(k) \, {\sf U}_S^{-1} &=  +  {\sf P}(k).\nonumber
\end{align}
for a corresponding unitary matrix ${\sf U}_S$. Chiral symmetry is only a nontrivial additional information if both $\sf T$ and $\sf C$ are broken.

Labelling the unit cell with index $j$ and the four sites within a unit cell with $\mu \in \{A,B,C,D\}$, the Lindblad generators describing the gain and loss processes for the dissipative topologically nontrivial case $g_1 = g_2 = g_0 \equiv g > 0$ discussed in the main text are
\begin{align}
    \hat L_{jA} &= \sqrt{gJ} \, \hat c_{j,A}^\dagger,\qquad \hat L_{jB} = \sqrt{gJ} \, \hat c_{j,B},\\
    \hat L_{jC} &= \sqrt{gJ} \, \hat c_{j,C},\qquad \hat L_{jD} = \sqrt{gJ} \, \hat c_{j,D}^\dagger.
\end{align}
%
For the overall loss, we have
\begin{equation}
    \hat L_{j\mu}^{(0)} = \sqrt{gJ} \, \hat c_{j,\mu}.
\end{equation}
With this, one finds the $8\times 8$ matrices
\begin{align}
    {\sf H}(k) &= \left[
    \begin{array}{cccc}
        0 & -\sigma_z & 0 & {\sf h}(k) \\
        -\sigma_z & 0 & -\sigma_z & 0 \\
        0 & -\sigma_z & 0 & -\sigma_z \\
        {\sf h}^*(k) & 0 & -\sigma_z & 0
    \end{array}
    \right],
\label{eqHk}
\end{align}
where we dropped the prefactor $J$. In eq.~\eqref{eqHk}, $\sigma_z$ denotes one of the Pauli matrices and
\begin{equation}
    {\sf h}(k) = \left[
    \begin{array}{cc}
        - e^{-4i k} & 0 \\
         0 & e^{4 ik}
    \end{array}
    \right].\nonumber
\end{equation}
Furthermore, for the matrices describing dissipation ($\sf D$) and fluctuations ($\sf P$), we obtain
\begin{align}
     {\sf D} = 
    \left[
    \begin{array}{cccc}
        \mathbf{1}_2 &  &  &  \\
         &   \mathbf{1}_2 &  &  \\
         &  &  \mathbf{1}_2 &  \\
         &  &  &  \mathbf{1}_2
    \end{array}
    \right], \ \ 
     {\sf P} &=  i \left[
    \begin{array}{cccc}
        \mathbf{0}_2 &  &  &  \\
         & \sigma_z &  &  \\
         &  & \sigma_z &  \\
         &  &  &  \mathbf{0}_2
    \end{array}
    \right],
\end{align}
where we dropped the common prefactor $\sqrt{gJ}$. Note that the dissipative topologically trivial case $g_1 = g_0 \equiv g>0$, and $g_2=-g$ just corresponds to an exchange of rows and columns in the matrix $\sf P$.

We find that $\sf H$, $\sf D$, and $\sf P$ possess time-reversal, charge conjugation and thus also chiral symmetry, according to eqs.~\eqref{eq:T}-\eqref{eq:S}. The corresponding unitary matrices read
\begin{equation}
     {\sf U}_T = \left[
    \begin{array}{cccc}
        \mathbf{1}_2 &  &  &  \\
         & -\mathbf{1}_2 & &  \\
         &  & \mathbf{1}_2 &  \\
         &  &  & -\mathbf{1}_2
    \end{array}
    \right],
\end{equation}
and
\begin{equation}
     {\sf U}_C = \left[
    \begin{array}{cccc}
         &  &  &  \sigma_x\\
         &  & \sigma_x &  \\
         & \sigma_x &  &  \\
         \sigma_x &  &  & 
    \end{array}
    \right],
\end{equation}
and ${\sf U}_S= {\sf U}_T\cdot {\sf U}_C$. This situation corresponds to the symmetry class BDI, which according to the periodic table of topological insulators allows in one spatial dimension for topologically distinct phases characterized by a $\mathbb{Z}$-valued index~\cite{Ryu:2010,Kitaev:2009}. 

\js{To complete our theoretical discussion of the topological system, we note that while there is not yet a general understanding of non-equilibrium invariants for density matrices, a topological number for the non-Hermitian 4-site lattice system has been identified theoretically as a normalized global Berry phase 
\begin{equation}
W=\sum_j \frac{i}{4\pi} \oint dk \langle \langle \psi_{j}|\partial_k|\psi_{j}\rangle    
\end{equation}
with lattice site index $j=1,...,4$ in the unit cell~\cite{Liang:2013,Takata:2018}. The left and right eigenstates fulfill $\hat H_k|\psi_j\rangle = \omega_k |\psi_j\rangle $ and $\hat H_k^\dagger|\psi_j\rangle\rangle=\omega^*_k|\psi_j\rangle \rangle$ and form a biorthonormal basis $(\{|\psi_j\rangle \},\{ |\psi_j\rangle\rangle \})$ that enables the extraction of geometric phases from non-Hermitian Hamiltonians. For the topological configurations (II) or (III) studied in the main text, the topological winding number takes discrete values $W=0$ or $1$ for $g_1g_2<0$ or $g_1g_2>0$, respectively.}

\section{Numerical simulations}
To validate the experimental results, we perform numerical simulations of the SPP dynamics in the waveguide arrays, which yield both the real-space evolution and the momentum-resolved spectral distributions. The numerical method is based on coupled mode theory~\cite{Pierce:1954}. The evolution is governed by the differential equation 
\begin{equation}
\frac{da_j}{dz}=i C_{j-1,j}a_{j-1}+i\beta_j a_j+i C_{j,j+1}a_{j+1},
\end{equation} 
where $a_j\equiv a_j(z)$ denotes the field amplitude in the $j$-th waveguide at position $z$. The spatially uniform coupling constants $C_{j,j\pm 1}=C$ between neighboring waveguides $j$, $j\pm 1$ are independent of $z$ and have been measured in the fabricated waveguide structures, see the plot in Fig.~1(c). The propagation constants $\beta_j=\mathrm{Re}\beta_j+i\ \mathrm{Im}\beta_j$ are composed of a globally constant $\mathrm{Re}\beta_j=\SI{6.6}{\micro\meter}^{-1}$ and a $j$-dependent $\mathrm{Im}\beta_j$; see Fig.~1(c) for its dependence on different chromium widths. For all simulations, the initial conditions are chosen such that the field at $z=0$ is concentrated in the first waveguide of a unit cell, either (i) in the bulk, (ii) at the sample edge, or (iii) at the interface between distinct topological domains. All simulations are performed for system sizes of 40 waveguides and $g_1=|g_2|$.

{\itshape Bulk excitation.} We simulate the SPP evolution following an excitation of a waveguide in the bulk of an array for two configurations, with and without losses, as shown in Fig.~2(b). While the lossless case realizes phase (I), the loss pattern in the dissipative case is tailored according to the topologically trivial phase (II), which is equivalent to the bulk of the topologically nontrivial phase (III). In phase (I), the simulation shows a conical spread of the wave packet, resembling a two-state quantum random walk with a $\cos$-shaped band structure in momentum space~\cite{Bleckmann:2017}. In phase (II) and (III), the population oscillates between the excited lossless waveguide and its lossless neighbor. For increased losses, the oscillation period becomes larger and the transverse transport into neighboring unit cells is suppressed. For $|g_2|>1$, the oscillation can be interpreted as a beating between the eigenstates above and below the band gap with a period of $2\pi/\Delta k_z$, where $\Delta k_z$ is the difference between the mean energies of the upper and lower bands visible in Fig.~1(b). The momentum-resolved spectrum for the topologically trivial phase (II) features population in the upper and lower energy bands, respectively. Due to dissipation, the main contribution to the SPP evolution occurs during the first half-period in the downward $-x$ direction; accordingly, the band structure appears slightly distorted and the band gap is clearly pronounced only in a limited momentum range around $k_x\approx \SI{-1}{\micro\meter}^{-1}$, as visible in the  middle column of Fig.~2(c).

{\itshape Edge excitation.} To study the topological edge mode in phase (III), we excite the outermost waveguide ({\itshape i.e.}, at the interface to the vacuum), where the expected probability density for the midgap state is largest. While for the lossless case the wave packet spreads entirely into the bulk, for a topologically nontrivial phase-(III) configuration the majority of the population remains localized in the excited waveguide and only a small amount spreads into the bulk. Accordingly, a flat band is visible in the momentum-resolved spectrum indicating the zero-energy topological mode residing in the band gap, see Fig.~2(c) (right column). For a phase-(II) configuration, the lossy waveguides can be either the last or the first two ones within a unit cell. Exciting the high-loss waveguide leads to a rapid population decrease and a broadened spectral distribution. If the lossless waveguide is excited, the intensity oscillates between the two outermost waveguides. Crucially, the states that contribute to this oscillatory dynamics are not topologically protected; neither resides in the band gap, exhibits an exponential decay towards the bulk, or features a finite topological winding number~\cite{Liang:2013,Takata:2018}, which here takes the value $W=0$.

\begin{figure}[t]
    \centering
    \includegraphics[width=1.0\columnwidth]{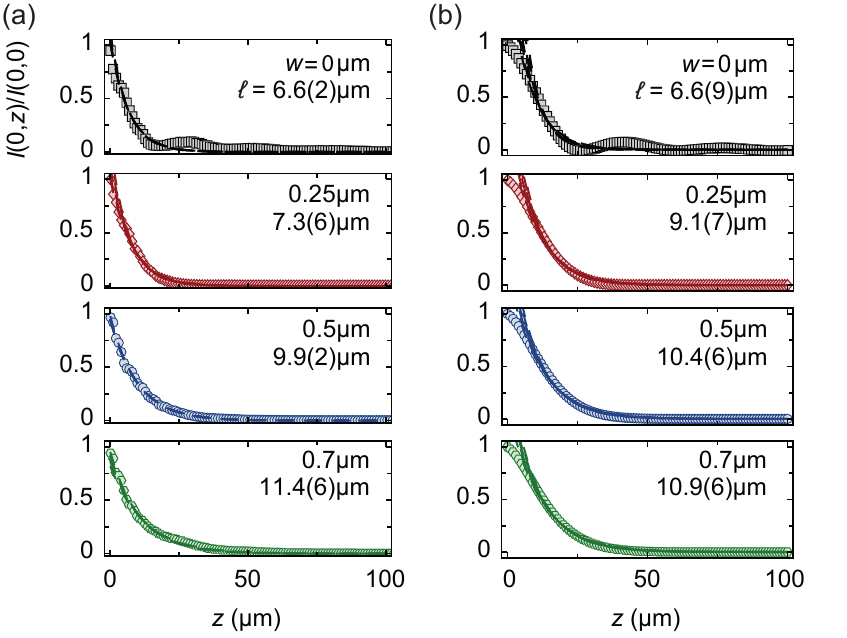}
\caption{\js{Line profiles of normalized intensity $I(x=0,z)/I(0,0)$ evolution in excited waveguide at interface between phase (II) and (III) [see Fig.~3(a) of the main text], for increasing losses from Chromium stripes of width $w=\{0,0.25,0.5,0.7\}~\SI{}{\micro\meter}$. (a)~Experimental and (b)~numerical data along with exponential fit $f(z)=a_0 \exp({-z/\ell})$ (dashed lines) over variable fit ranges $[0{-}5:27]~\SI{}{\micro\meter}$ for experiment and $[4{-}10:80]~\SI{}{\micro\meter}$ for numerics, respectively. The error on the decay length $\ell$ is calculated from the standard deviation of the fit parameters for the different fit ranges.}}
    \label{fig:SI2}
\end{figure}

{\itshape Interface excitation.} To simulate the emergence of a topologically protected edge mode at a 'nontrivial' ({\itshape i.e.}, not with the vacuum) interface between distinct topological domains in phases (II) and (III), respectively, the first waveguide of phase (III) unit cell is excited. For weak absorption $\mathrm{Im}\beta$, the simulation shows a diffusion of the population into the neighboring unit cells. For larger $\mathrm{Im}\beta$, diffusion into the bulk is suppressed and a majority of the population remains at the interface waveguide for longer propagation distances, see Fig.~3(c). The spread into the bulk is understood from considering the eigenstates localized at the interface. For the interface modes, the intensity on the 4th waveguide above or below the interface ({\itshape i.e.}, shifted by one unit cell) is larger as compared to the neighbors, similar to the topological edge mode. Increasing absorption leads to an enhanced intensity at the interface compared to the neighboring unit cells. For the largest simulated $\mathrm{Im}\beta$, the population spread is limited to the neighboring unit cell and, therefore, simulating two unit cells would be sufficient.

\section{Intensity decay at topological interface}
\js{Figure~\ref{fig:SI2} shows line profiles of the normalized intensity in the excited waveguide at an interface between phases (II) and (III) for increasing dissipation, as realized by increasing width $w=\{0,0.25,0.5,0.7\}~\SI{}{\micro\meter}$ of the absorbing Chromium width. The experimental traces in Fig.~\ref{fig:SI2}(a) are extracted from the real-space intensity distribution shown in Fig.~3(a) of the main text, while the traces in Fig.~\ref{fig:SI2}(b) give the corresponding numerically simulated results. Both data sets are fitted with an exponentially decaying function, yielding the decay length $\ell$. The error of $\ell$ is estimated by fitting different $z$ ranges, either including or excluding the initial part of the intensity evolution in the waveguide that deviates from a pure clear exponential decay. The fit results are rather insensitive on the choice of the upper end of the fitting range. Without losses $w=\SI{0}{\micro\meter}$, on the one hand, the visible intensity oscillations are a consequence of the interference pattern of the random walk of the particle in the lattice. In a waveguide lattice with spatially distributed losses, on the other hand, the oscillations are suppressed and the intensity evolution is well described by an exponential decay already after a few $\SI{}{\micro\meter}$ of propagation. The increased decay length based on the exponential fit with larger $w$ indicates that the topological interface state enhances the propagation distance in the excited waveguide between the two distinct phases (II) and (III).}


\end{document}